\documentclass[journal,10pt]{IEEEtran}
\pdfoutput=1
\normalsize

\ifCLASSINFOpdf
 \usepackage[pdftex]{graphicx}
\else
 \usepackage[dvips]{graphicx}
\fi

\DeclareGraphicsExtensions{.eps}
\usepackage[cmex10]{amsmath}
\usepackage{bm}
\usepackage{upgreek}
\usepackage{epsfig}
\usepackage[noadjust]{cite}
\usepackage{latexsym}
\usepackage{multirow}
\usepackage{epstopdf}
\usepackage{color}  
\usepackage{amssymb}
\graphicspath{{../Figures/}}
\usepackage{color}
\usepackage{bbm}
\usepackage{mathtools}
\usepackage{url}

\usepackage{cite}
\usepackage[tight,footnotesize]{subfigure}
\usepackage{balance}

\usepackage{color, soul}

\usepackage{siunitx}
\DeclareSIUnit\bps{bps}
\DeclareSIUnit\Torr{Torr}
\DeclareSIUnit\torr{Torr}
\DeclareSIUnit\sample{Sa}

\usepackage{cite}
\usepackage{balance}

\begin{document}

\title{Terahertz Communications (TeraCom): Challenges and Impact on 6G Wireless Systems}
\author{
        Chong~Han,~\IEEEmembership{Member,~IEEE,}
        Yongzhi Wu,~\IEEEmembership{Student Member,~IEEE,}
         Zhi~Chen,~\IEEEmembership{Member,~IEEE,}
         and Xudong~Wang,~\IEEEmembership{Fellow,~IEEE}
        
\thanks{Chong Han, Yongzhi Wu and Xudong Wang are with Shanghai Jiao Tong University, Shanghai, China (email: \{chong.han, yongzhi.wu, wxudong\}@sjtu.edu.cn).}
\thanks{Zhi Chen is with University of Science and Technology of China, China (email: chenzhi@uestc.edu.cn).}
}

%
%

{}
%



\maketitle

\begin{abstract}
Terahertz communications are envisioned as a key technology for 6G, which requires 100+ Gbps data rates, 1-millisecond latency, among other performance metrics. 
As a fundamental wireless infrastructure, the THz communication can boost abundant promising applications, including next-generation WLAN systems like Tera-WiFi, THz wireless backhaul, as well as other long-awaited novel communication paradigms. 
Serving as a basis of efficient wireless communication and networking design, this paper highlights the key THz channel features and recent advancements in device technologies.
In light of these, impact and guidelines on 6G wireless communication and networking are elaborated.
We believe the progress of THz technologies is helping finally close the so called THz Gap, and will realize THz communications as a pillar of 6G wireless systems. 
\end{abstract}


%
\IEEEpeerreviewmaketitle

\section{Introduction}
%
%
%
%

\begin{figure*}[t]
	\centering
	\includegraphics[width=0.95\textwidth]{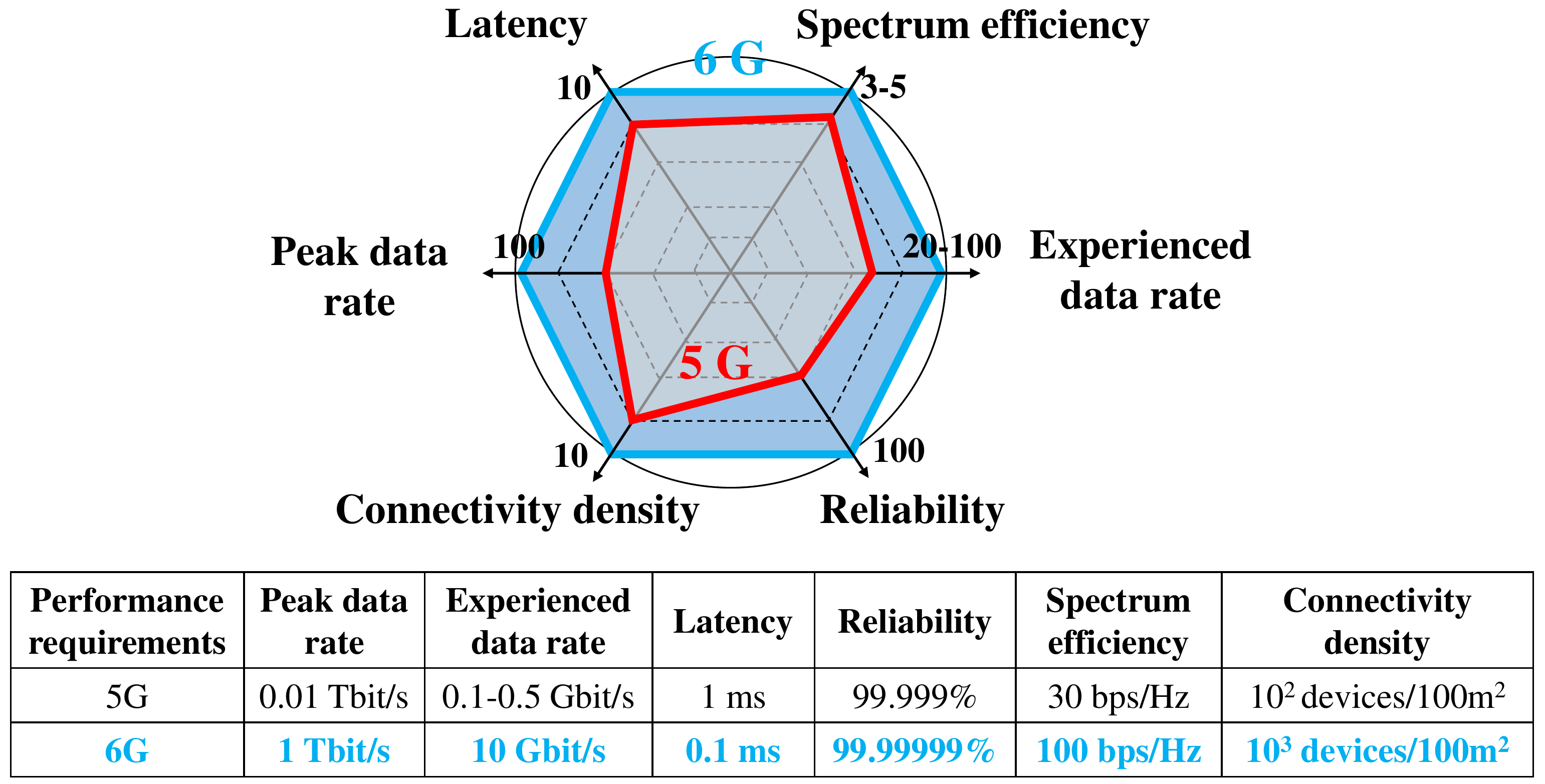}
	\caption{Performance requirements for 6G, as a disruptive step forward from 5G.}\label{fig:6G_views}
\end{figure*}

\IEEEPARstart{I}{n} current wireless communication systems, revolutionary enhancement of data transmission rate is witnessed, as a result of the way of creation, sharing and consumption of innovative information and communication technologies. On one hand, the peak data rate is expected to reach 10 Gbps in 5G and continue boosting.
On the other hand, thanks to the immersion of the Internet of Things (IoT) paradigm, ubiquitous wireless connectivity that connects with more than 11 billion devices is foreseen by 2020.
Following this trend, Terabit-per-second (Tbps) links are expected to become a reality within the next decade~\cite{akyildiz2018combating}.
As summarized in Fig.~\ref{fig:6G_views}, moving forward from 5G, the performance of 6G includes the following six aspects of disruptive improvement, namely, (i) 1 Tbps peak data rates, (ii) up to 10 Gbps experienced data rates for users, (iii) shorter than 0.1 ms latency, (iv) 100 bps/Hz spectrum efficiency, (v) 99.99999$\%$ reliability, and (vi) enormous connectivity density of more than 10\textsuperscript{3} devices per 100 m\textsuperscript{2} \cite{Zhang2019sixG}.

One critical question arises naturally: \textit{can the current frequency spectrum satisfy these demands?} The answer from the authors is NO!
Millimeter-wave (mm-wave) communications (30-300~GHz) have been officially adopted in recent 5G cellular systems, confirmed by the allocation of several mm-wave sub-band for licensed communication by the Federal Communication Commission (FCC) \cite{Rappaport2019wireless}.
While the trend for higher carrier frequencies is apparent, it is still difficult for mm-wave systems to support Tbps data rates, limited by the total consecutive available bandwidth of less than 10 GHz in the mm-wave systems.
Neighboring to the mm-wave spectrum, the \textit{Terahertz (THz) band (0.1-10~THz)} is revealing its potential as a key wireless technology to fulfill the future demands for \textbf{6G wireless systems}, thanks to its four strengths: 
1) tens up to hundred GHz bandwidth resource, 2) pico-second-level symbol duration, 3) integration of thousands of sub-millimeter-long antennas, 4) weak interference without full legacy regulation.
Being known as the THz gap for many years, the THz band has been one of the least explored frequency bands in the electromagnetic (EM) spectrum, for lack of efficient THz transceivers and antennas. Nevertheless, practical THz communication systems are enabled by the major progress in the last ten years~\cite{crowe2017terahertz}.

\begin{figure*}[!t]
    \centering
    \includegraphics[width=0.9\textwidth]{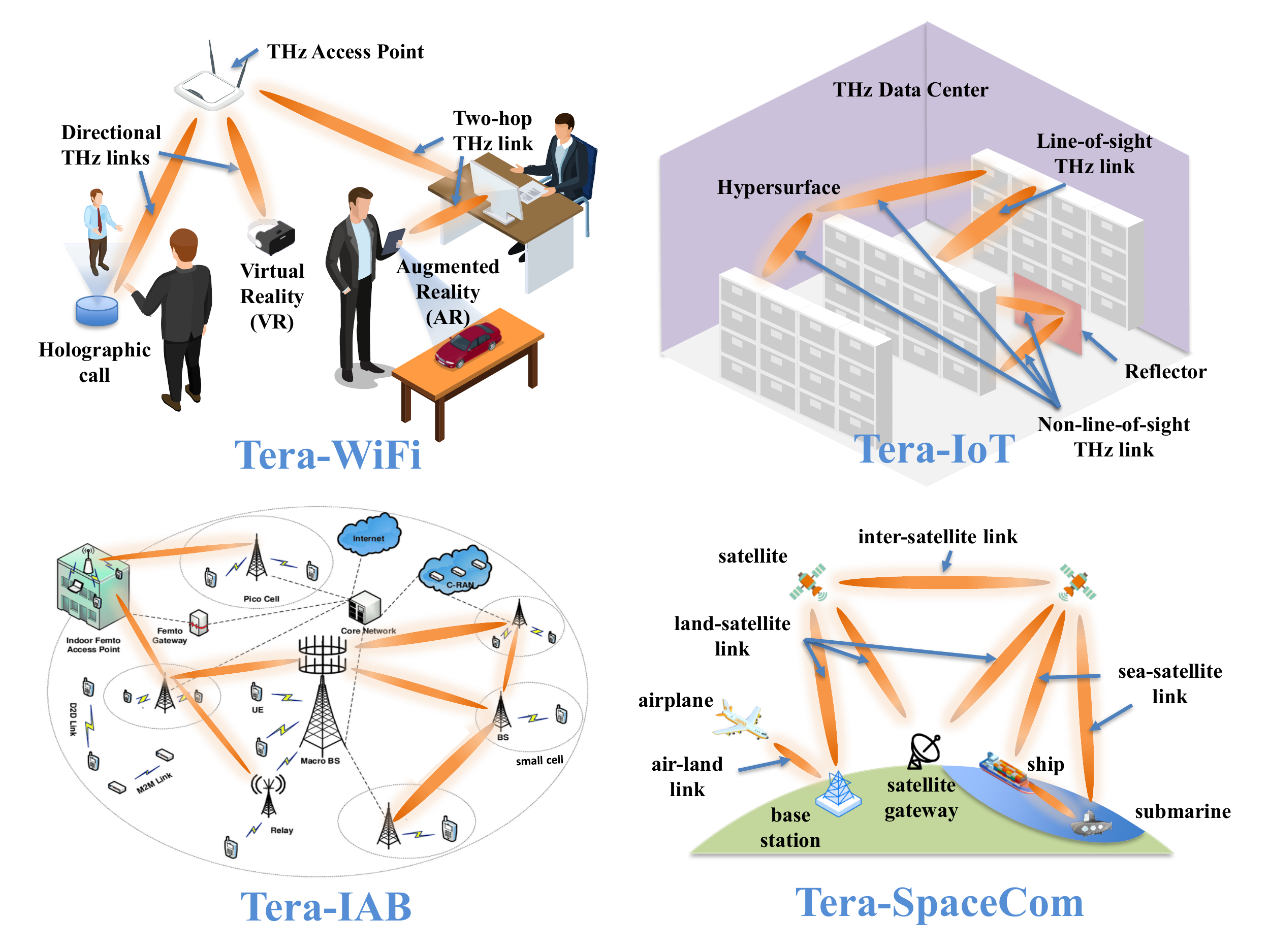}
    \caption{6G applications operating in the THz band, including Tbps WLAN system (\textit{Tera-WiFi}), Tbps Internet-of-Things (\textit{Tera-IoT}) in wireless data center, Tbps integrated access and backhaul (\textit{Tera-IAB}) wireless networks, and ultra-broadband THz space communications (\textit{Tera-SpaceCom}).}
    \label{fig:applications}
\end{figure*}

The THz spectrum can resolve the spectrum scarcity problem and tremendously enhance current wireless system capacity~\cite{Petrov2018last}. Various promising applications are envisaged, such as Tbps WLAN system (\textit{Tera-WiFi}), Tbps Internet-of-Things (\textit{Tera-IoT}) in wireless data center, Tbps integrated access backhaul (\textit{Tera-IAB}) wireless networks, and ultra-broadband THz space communications (\textit{Tera-SpaceCom}), as illustrated in Fig.~\ref{fig:applications}.
Besides these macro/micro-scale applications, the THz band can be utilized for wireless connections in nanomachine networks, to enable wireless networks-on-chip communications (WiNoC) and the Internet of Nano-Things (IoNT)~\cite{8663550}, motivated by the state-of-the-art nanoscale transceivers and antennas that oscillate in the THz band.


Challenges exist with the promising opportunities at the same time, particularly from the very high propagation loss and the constrained communication distance. 
Therefore, the road to close the THz gap and realize the THz communications needs to be paved by the communication community jointly.
In this direction, global research activities are ongoing, including ICT-09-2017 cluster funded by Europe Horizon 2020, key programs funded by Chinese Ministry of Science and Technology, and multiple ongoing NSF grants in the USA, among others. 
The first wireless communications standard, IEEE 802.15.3d (WPAN), is published in 2017, which operates at the 300~GHz frequency range to support 100 Gbps and above wireless links.
Since 2019, the first and second international workshops focused on Terahertz communications have been successfully held, while the momentum continues globally.

In this article we highlight the unique features of THz channel and device technologies.
In light of these, we further delineate the impact as well as guidelines on 6G THz wireless communications and networking, from the perspectives of ultra-massive MIMO and razor-sharp beam, modulation and coding, medium access control, routing and THz mesh networking protocols.
Finally, we conclude the article.

\section{Terahertz Band vs. Microwave and Millimeter Wave vs. Free-space Optics: Key Differences}\label{sec:device}
In this section, we demonstrate side-by-side comparison among THz band, microwave and millimeter wave, and free-space optics (FSO), from the perspectives of propagation characteristics and device technologies.

\subsection{Channel and Propagation Characteristics}
\label{sec:channel}
As the key to understanding the THz spectrum and laying out fundamentals for communications and networking, THz channel modeling and characterization have drawn attention in the past several years, and a comprehensive survey on THz propagation modeling is in~\cite{han2018propagation}.
Instead of elaborating the channel modeling strategies, here we focus on the fundamental spectrum and channel characteristics, as well as the potential impact on 6G wireless communication and networking, which are summarized in Table~\ref{tab:overview}.
The foremost propagation property is free-space path loss, which consists of spreading loss as well as atmospheric loss (also known as molecular absorption loss).
Moreover, diffuse scattering and specular reflection, diffraction and shadowing, weather influences, as well as scintillation effects are elaborated.

\textit{Spreading Loss:} On one hand, the spreading loss increases quadratically with frequency, if we consider antenna aperture shrinks as frequency increases at both link ends while antenna gains are frequency-independent, as defined by the Friis' law.
On the other hand, this spreading loss decreases with square of frequency, when considering antennas have a constant physical area and frequency-dependent gains at both link ends.
Generally the former case is adopted and numerically, the free space path loss of a 10-meter THz wireless link reaches over  100~dB \cite{Guan2019measurement}, which makes long-range communications challenging.

\textit{Atmospheric Loss:}
The molecular absorption loss is caused due to polar gas molecules in the propagation medium absorb THz wave energies for their rotational transition energies. Both spreading and atmospheric effects attenuate THz wave propagation.
Mainly triggered by water vapor and oxygen molecules at THz frequencies, the absorption peaks create spectral windows and make 1-10~THz spectrum strongly opaque.
These spectral windows span over multi-GHz bandwidths, which are strongly dependent on the distance~\cite{han2018propagation}.
The atmospheric loss is not observed at microwave frequencies.
However, this loss is observed, owing to the oxygen molecules at the millimeter wave, and water vapor and carbon dioxide molecules at FSO, respectively.


\textit{Diffuse Scattering and Specular Reflection:}
When the EM wave impinges on rough surfaces, diffuse scattering is created where scattered rays travel to various forward and backward directions. 
This is due to the fact that as the wavelength reduces to millimeter and sub-millimeter at THz frequencies, which is comparable with the surface roughness of common objects 
Among all the directions, specular reflection refers to the scattered ray along the identical angle as the incident angle.
The resulting specular reflection loss depends on the electrical thickness of the object surface which depends on frequency.
More specifically, both the height standard deviation and correlation length of a rough surface affect the power concentrated along the specular direction.
On one hand, the scattered power level is more influenced by the rms height parameter, which decreases with frequency.
On the other hand, the scattered beam width is more related to the surface correlation parameter, which increases with frequency.
In a multi-path propagation model, the specular reflection loss is smaller than the diffuse scattering loss in general, and contributes more significantly to the received power.
However when frequency is very large and roughness is serious, rough surface diffusely radiates uniformly into all directions.

\textit{Diffraction, Shadowing and Line-of-Sight Probability:}
With the increase of frequency, the diffraction efficiency reduces due to the fact that the sharp shadows are brought by building walls or people and other obstacles. In practice, the diffraction path can generally be ignored unless the receiver is located in a very closed region near the incident shadow boundary.
Moreover, shadow fading variance becomes more pronounced with a higher frequency, due to the larger diffraction loss.
To understand the influence of shadowing,
the size of the first Fresnel zone, i.e., the area particularly vulnerable to shadowing objects, decreases with the square root of the wavelength.
Furthermore, the line-of-sight (LoS) probability is not necessarily dependent on the frequency.
However, when EM wave encounters an object, transmission power through it decreases almost uniformly with the increase of frequency, since these exist skin effect in lossy medium.

\textit{Weather Influences:}
THz wave propagation is further attenuated by weather influences, or more specifically, airborne particulates from rain, fog, smoke, among others.
This effect can be modeled by either the Mie scattering theory when the size of airborne particles is comparable to the wavelength, or the Rayleigh approximation to Mie scattering when the size of airborne particles is smaller than the wavelength.
With even smaller wavelengths, the infra-red or free-space optics are strongly attenuated in non-clear sky weather. By contrast,  THz waves can propagate through small airborne particulates and are better immune to weather influences.

\textit{Scintillation Effects:}
Owing to thermal and turbulences near the ground, inhomogeneities of temperature and pressure across time and space in the air can produce scintillation.
Being observed as twinkling, the beam cross-section on the receiver side appears as a speckle pattern that contains severe temporal variations on intensity.
This is caused by the removal of flat phase front of the beam, for the two reasons as follows.
First, the wave front of a beam involves  variation of  refractive index. Second, the wave front is further influenced by unevenly distributed  deflection and interference.
As a result, the maximum transmission distance of free-space optical communication links is mainly limited by the effects of scintillation, while THz wave is much less susceptible to it.

\begin{table*}
\centering
\caption{THz channel features and impact on 6G wireless communication and networking.}
\label{tab:overview}
\begin{tabular}{|c|c|c|c|}
\hline
 \textbf{Parameter}      & \textbf{Dependence on frequency}  & \textbf{Impact on 6G THz systems} & \textbf{THz vs. Microwave and FSO}        \\ \hline
Spreading Loss & \begin{tabular}[c]{@{}c@{}} Quadratic increase with \\ decreasing area and constant gains; \\ Quadratic decrease with constant \\area and frequency-dependent gains \end{tabular}    & Distance limitation& \begin{tabular}[c]{@{}c@{}}Higher than microwave, \\ lower than FSO \end{tabular} \\ \hline
Atmospheric Loss &\begin{tabular}[c]{@{}c@{}}Frequency-dependent \\ path loss peaks appear \end{tabular}& \begin{tabular}[c]{@{}c@{}}Frequency-dependent spectral \\ windows with varying bandwidth\end{tabular}  & \begin{tabular}[c]{@{}c@{}} No clear effect at microwave frequencies, \\ oxygen molecules at millimeter wave, \\ water and oxygen molecules at THz,\\ water and carbon dioxide molecules at FSO\end{tabular} \\ \hline
\begin{tabular}[c]{@{}c@{}}Diffuse Scattering \\ and Specular Reflection\end{tabular}   &\begin{tabular}[c]{@{}c@{}} Diffuse scattering increases \\ with frequency. Specular reflection loss \\is frequency-dependent\end{tabular}& Limited multi-path and high sparsity &  \begin{tabular}[c]{@{}c@{}}Stronger than microwave,\\ weaker than FSO\end{tabular} \\ \hline
\begin{tabular}[c]{@{}c@{}} Diffraction, Shadowing \\and LoS Probability \end{tabular}& \begin{tabular}[c]{@{}c@{}}Negligible diffraction. Shadowing and \\penetration losses increase with frequency. \\ Frequency-independent LoS probability \end{tabular} &  \begin{tabular}[c]{@{}c@{}}Limited multi-path, high sparsity\\ and dense spatial reuse\end{tabular} & \begin{tabular}[c]{@{}c@{}}Stronger than microwave,\\ weaker than FSO\end{tabular} \\ \hline
Weather Influences & \begin{tabular}[c]{@{}c@{}}Frequency-dependent airborne\\ particulates scattering \end{tabular}& \begin{tabular}[c]{@{}c@{}}Potential constraint in \\ THz outdoor  communications \\ with heavy rain attenuation \end{tabular}& \begin{tabular}[c]{@{}c@{}}Stronger than microwave,\\ weaker than FSO\end{tabular}  \\ \hline
Scintillation Effects & Increase with frequency& \begin{tabular}[c]{@{}c@{}}Constraint in THz \\ space communications \end{tabular}& \begin{tabular}[c]{@{}c@{}}No clear effects at microwave, \\THz is less susceptible than FSO\end{tabular}\\ \hline
\end{tabular}
\end{table*}

\subsection{Device Technologies}
\subsubsection{Terahertz Transceivers}
\label{sec:transceivers}

Although Terahertz sensing has been extensively studied since the 1990s, the lack of compact high-power signal transmitters and high-sensitivity detectors working at room temperature has been a major challenge for THz communications for many decades.
Many recent advancements following different technology paths are jointly filling the so-called THz gap.  

On the one hand, \textit{electronic technologies}~\cite{crowe2017terahertz} including standard silicon CMOS technology, silicon-germanium BiCMOS technology, and III-V semiconductor-based High Electron Mobility Transistor (HEMT), metamorphic HEMT, Heterojunction Biopolar Transistor (HBT) and Schottky diode technology have been vastly advanced and can be found at the basis of sources, amplifiers and mixers able to operate at frequencies close to \SI{1}{\tera\hertz}. With these technologies, THz signals are principally generated by means of up-conversion from lower frequencies through chains of frequency multipliers.
The achievable transmission power ranges from hundreds of milliWatts (mWs) at 240~GHz to a few mWs at 1~THz.

On the other hand, \textit{photonic technologies}~\cite{nagatsuma2016advances}, including optical down-conversion systems based on photomixers or photoconductive antennas, uni-traveling carrier photodiodes (UTC) and quantum cascade lasers (QCLs), have also been demonstrated as potential enablers of practical THz communication systems. While the power of optical systems is much lower than that of electronic systems, the very high speed at which photonic signals can be modulated and processed plays to their advantage. Moreover, hybrid combinations are possible too, in which the transmitter and the receiver are based on electronic and photonic technologies, respectively. Nevertheless, the complexity in heterogeneous integration of electronic \& photonic technologies is an aspect that needs to be taken into account in this approach~\cite{sengupta2018terahertz}.

\begin{table*}
\centering
\caption{Summary of hardware performance of THz vs. microwave and FSO systems~\cite{8663550,nagatsuma2016advances, sengupta2018terahertz}.}
\label{tab:device}
\begin{tabular}{|c|c|c|c|}
\hline
 \textbf{Parameter}      & \textbf{Microwave}  & \textbf{Terahertz} & \textbf{FSO}        \\ \hline
Bandwidth & MHz up to several GHz    & Tens up to hundreds of GHz & Tens of GHz  \\ \hline
Transmitter power & Watts    & Milliwatts & Dozens of Watts  \\ \hline
Receiver noise figure & About 5 dB  & About 15 dB & About 15 dB \\ \hline
Receiver sensitivity & -120 dBm  &-48 dBm & -56 dBm \\ \hline
Signal processing speed & On the order of Gbps  & On the order of 10 Gbps & On the order of 10 Gbps \\ \hline
Data rate & At most 10 Gbps & On the order of 100 Gbps & On the order of 100 Gbps \\ \hline
System size & Large & Small & Bulky \\ \hline
Power consumption & High & Low & High \\ \hline
Communication continuity & All day & All day & Intermittent \\ \hline
\end{tabular}
\end{table*}


Different from the abovementioned electronic or photonic technologies, the recent adoption of nanomaterials has opened up a new door to develop novel plasmonic devices for THz communications, for instance, using graphene.
These devices are intrinsically small, efficiently operate at THz frequencies, and can support very large communication bandwidths. Contrary to electronic and optical technologies, which rely on the up-conversion of microwave and mm-wave signals or down-conversion of optical signals, the direct generation of THz signals is possible in hybrid graphene/III-V semiconductor devices. While still at an earlier stage, this technology can drastically increase the efficiency of THz communications systems thanks to the lack of energy loss through harmonics. 

A summary on the hardware performance of THz devices in contrast with microwave and FSO is presented in Table~\ref{tab:device}~\cite{8663550,nagatsuma2016advances,sengupta2018terahertz}.

\subsubsection{Terahertz Antennas and Arrays}
\label{sec:antennas}

The low transmission power of THz transceivers motivates the use of directional antennas. Several traditional antenna designs, including diagonal horn antennas, Cassegrain reflector antennas, and lens antennas are commercially available at frequencies under 1 THz. The small wavelength of THz signals (from 3 millimeters at 100 GHz to 30~$\mu$m at 10~THz) allows for these antennas to be very small. This property also allows more innovative designs, including multi-reflector antennas and lens-integrated antennas, all in very small footprints.

In addition, similarly as in the case of THz transceivers, new nanomaterials including can be utilized to design fundamentally new types of antennas.
Similar to the tranceiver design, graphene can be further employed to produce plasmonic nano-antennas radiating THz signals~\cite{jornet2013graphene}.
Nano-antennas leverage the propagation properties of Surface Plasmon Polariton (SPP) waves, i.e., confined electromagnetic waves that appear at the interface of a metal and a dielectric, on graphene, to efficiently radiate at THz frequencies.
Compared to traditional metallic antennas, they are two orders of magnitude smaller, which allows them to be integrated in virtually anything and, in addition, their frequency response can be electronically tuned.


\subsubsection{Terahertz Reconfigurable Intelligent Surfaces (RIS)}
\label{sec:ris}
Besides using the antenna arrays in transmission and reception, novel reconfigurable intelligent surfaces (RIS), or equivalently hypersurfaces~\cite{liaskos2018new}, can be utilized to control the propagation of THz signals, customizing EM wave absorbing, reflecting, polarization and phase shifting, collimation and focusing, among others.
Compared to traditional reflectors or relays, RIS allows for customization of the laws of electromagnetic propagation precisely and dynamically, which is realized by a set of conductive meta-atoms and switch elements on a dielectric substrate~\cite{liaskos2018new}.


The idea of RIS was also proposed for microwave indoor communications. 
Since microwave has a strong diffraction ability and circumvents obstacles preeminently, it is unnecessary for microwave communications to deploy the RIS. Instead, the RIS is a promising choice for FSO system to overcome the LoS blockage. 
However, the implementation of optical RIS is challenging.
By contrast, THz RIS can effectively overcome the interruption of obstacles by steering waves to completely customized directions. In addition to mitigate the LoS blockage issue, the RIS can improve coverage by acting as relays, as well as security in THz systems, by directing links to surpass eavesdroppers.

\section{Impact and Guidelines on 6G Wireless Communication and Networking}
\label{sec:communication}
We describe next the impact and guidelines on 6G wireless communication and networking,
from the perspectives of ultra-massive MIMO (UM-MIMO) and razor-sharp beam, modulation and coding, medium access  control,  routing  and  THz  mesh  networking  protocols.

\subsection{Ultra-Massive MIMO and Razor-Sharp Beamforming}
With sub-millimeter antennas, ultra-massive MIMO systems (1024$\times$1024) are promising for THz wireless communications~\cite{akyildiz2016ummimo}. 
How to group and control the very large antenna array need to be carefully treated in UM-MIMO systems, to achieve performance tradeoff among spectral efficiency, energy efficiency and distance enhancement.
On the one hand, antenna arrays can be summoned to steer highly directional ``razor-sharp'' radiation toward the strongest propagation path as beamforming, using analog, digital and hybrid architectures.
On the other hand, with a combination of multiple data streams, multiple RF chains and spatial degree of freedom of the THz channel with high sparsity, UM-MIMO can enable spatial multiplexing to increase the capacity, in conjunction with beamforming.

\begin{figure*}[t]
\centering
        \includegraphics[width=0.8\textwidth]{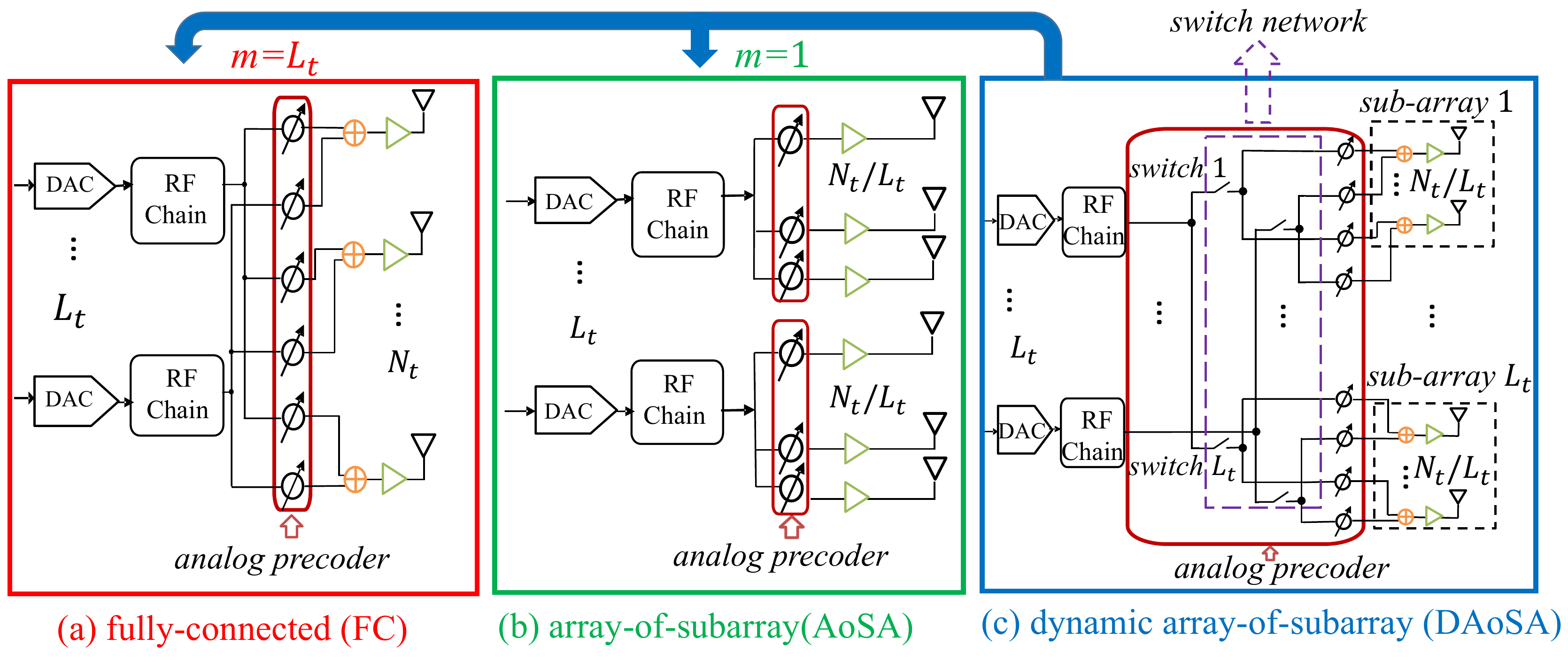} 
         \caption{Three hybrid beamforming architectures for THz communications, namely, (a) fully-connected (FC), (b) array-of-subarray (AoSA), and (c) dynamic AoSA, where $m$ denotes the number of subarrays that one RF chains\ connects to. When $m$ equals to the number of antennas $L_t$ or 1, DAoSA is equivalent to either FC or AoSA, respectively.} 
         \label{fig:beamforming}
 \end{figure*}

Among the three distinctive beamforming architectures, the hybrid one is appealing to THz communications due to its balance among good spectrum efficiency, low hardware complexity and energy efficiency.
In the hybrid beamforming, signal processing is decoupled into a digital baseband and a radio-frequency part through a phase shifter network.
The hybrid beamforming can be further clarified into three types, namely, full-connected, array-of-subarray (AoSA), and dynamic AoSA architectures, depending on the connection relationship between the RF chains and antennas, as illustrated in Fig.~\ref{fig:beamforming}.
To achieve efficient THz UM-MIMO communications, spectral efficiency, energy efficiency, as well as hardware complexity need to be jointly considered.
Furthermore, the low-rank of the THz multi-antenna channel, i.e., sparsity, needs to be rigorously treated to exploit spatial multiplexing design, by accounting for the possible spherical propagation of THz signals over widely-spaced antenna arrays.

\subsection{Modulation and Coding}
Due to the distance-related bandwidth in the THz spectrum, proper modulation techniques need to be developed for different applications based on the targeted transmission distance.
In short-range communications within one meter, modulations based on the exchange of one-hundred-femtosecond-long pulses can be explored as pulse amplitude modulation (PAM), on-off keying (OOK), and pulse position modulation (PPM).
These very short THz pulses can be easily generated and detected with photonic and plasma wave devices. 
For long-range communications, transmission waveforms can be dynamically adapted in THz spectral windows that are  determined by molecular absorption. This results in  distance-adaptive multi-wideband pulse modulation. 
Furthermore, for THz secure and covert applications \cite{ma2018security}, spread spectrum modulation techniques can be explored, thanks to the very large available bandwidth. In addition to invoking the techniques of time-hopping and frequency-hopping, the unique THz absorption peaks created atmospheric attenuation can be explored to prevent a signal transmission from hostile eavesdropping.

The next generation of forward error correction (FEC) schemes needs to be designed and implemented to overcome the THz channel errors. 
In order to design effective error control policies, it is stringent to characterize the nature of channel errors, which are affected by stochastic models of noise, multi-path fading, and interference.
After understanding the error sources in THz wireless systems, advanced state-of-the-art such as turbo, low density parity check (LDPC), and polar codes can be evaluated as possible candidates.
Instead of adopting the existing coding schemes, disruptive types of ultra-low-complexity channel coding schemes can be developed for THz communications. 
The developed FEC technology needs to be validated and demonstrated in future THz chipset.

\subsection{Medium Access Control (MAC)}
Many challenges on THz MAC design that are posed by the features of THz communications are elaborated as follows.
First, serious deafness problem, caused by the directivity of razor-sharp beams, could increase difficulty of alignment between a transmitter and receiver.
Second, due to the requirement for the coverage of transmission and the demand of deafness avoidance, the network discovery and coupling process become complicated when selecting a control channel.
The selection control channel needs to decide the frequency band, i.e., either microwave or THz band, as well as the type of sending-reception antenna modes, i.e., omni-, semi-, or fully-directivity.
Third, the cell boundary, which is determined by the received signal strength (RSS), might become amorphous rather than the regular circular or hexagonal shapes, owing to the existence of various human or wall LoS blockage.
Last but not least, while the multi-user interference (MUI) in THz communications is reduced, it is still required to monitor interference and propose effective concurrent transmission scheduling, especially in dense networks with spatial reuse.

These challenges are addressed in the in-depth literature ~\cite{han2018medium}, which investigates MAC protocols in the THz wireless networks.
Exploring the THz distinctive spectrum features, the distance-dependent spectral windows can facilitate concurrent communications based on the principle of long-user-central-window (LUCW), i.e., in a THz window, the central sub-band is allocated to the long user while the side-sub-band is assigned to the short user, without creating interference.
Combined LUCW with successive interference cancellation (SIC) at the receiver side and beamforming technologies, a promising attempt on THz non-orthogonal multiple access (NOMA) has a great potential to jointly optimize the utilization of beam, bandwidth and power resources.


\subsection{THz Mesh Networking}
To extend communication distances and network coverage for various applications of THz communications, it is indispensable to consider mesh networking based on active relaying nodes, reflectors, and RIS. Since THz links are prone to breakdown, it is challenging to deliver stable end-to-end networking across multiple hops. To ensure stability in mesh networking, a key step is to conduct link maintenance and topology management proactively based on an efficient control signaling procedure via in-band signaling, dual-band operation, or both. The link maintenance procedure should be prompt enough to detect broken links, and then recover or bypass such broken links.
Fast updates of fresh link status should also be achieved by the topology management procedure.

\begin{figure}[t]
	\centering
	\subfigure[The aggregate throughput versus traffic load.]{\includegraphics[width=.4\textwidth]{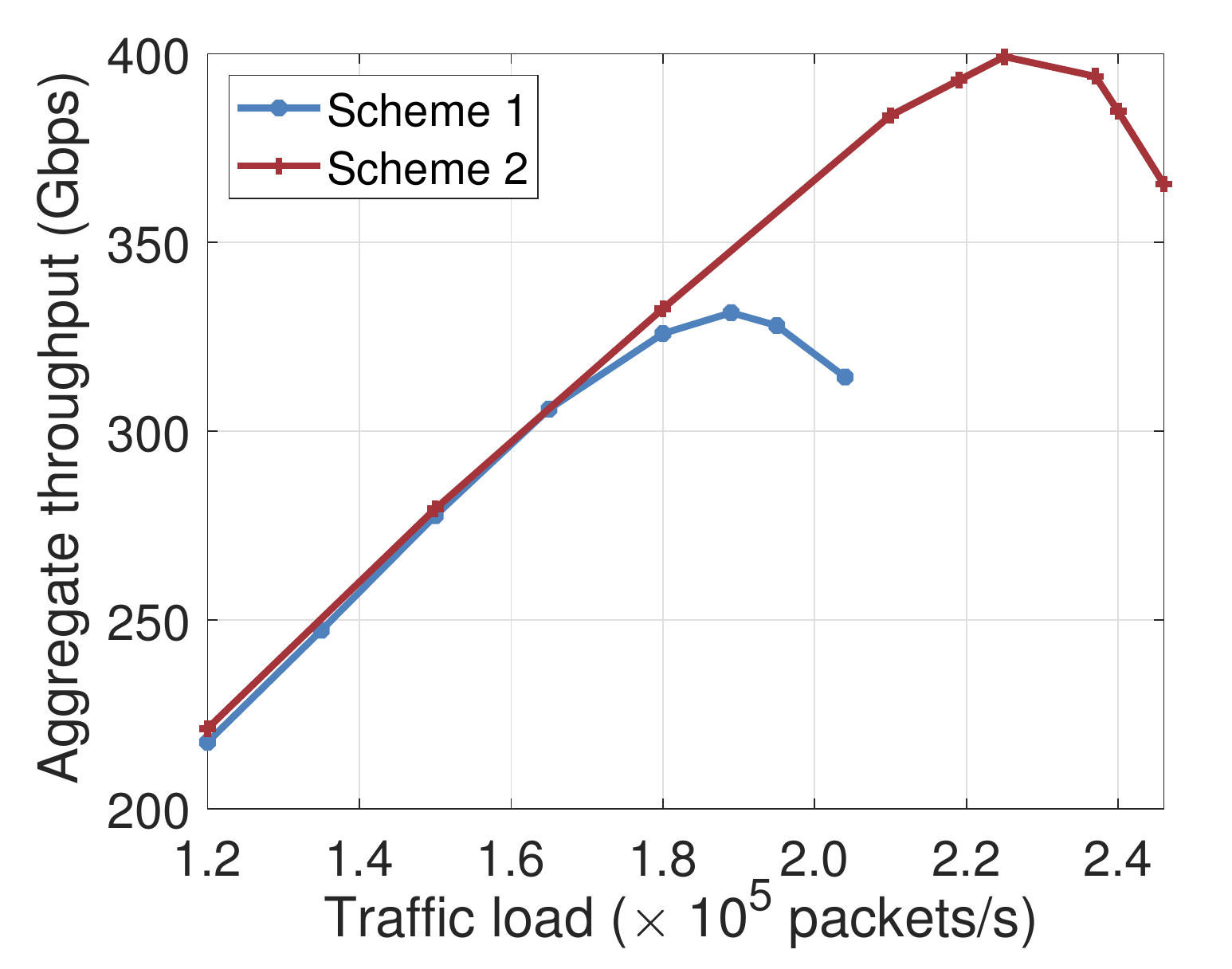}\label{fig:thru}}
	\subfigure[The end-to-end delay versus traffic load.]{\includegraphics[width=.4\textwidth]{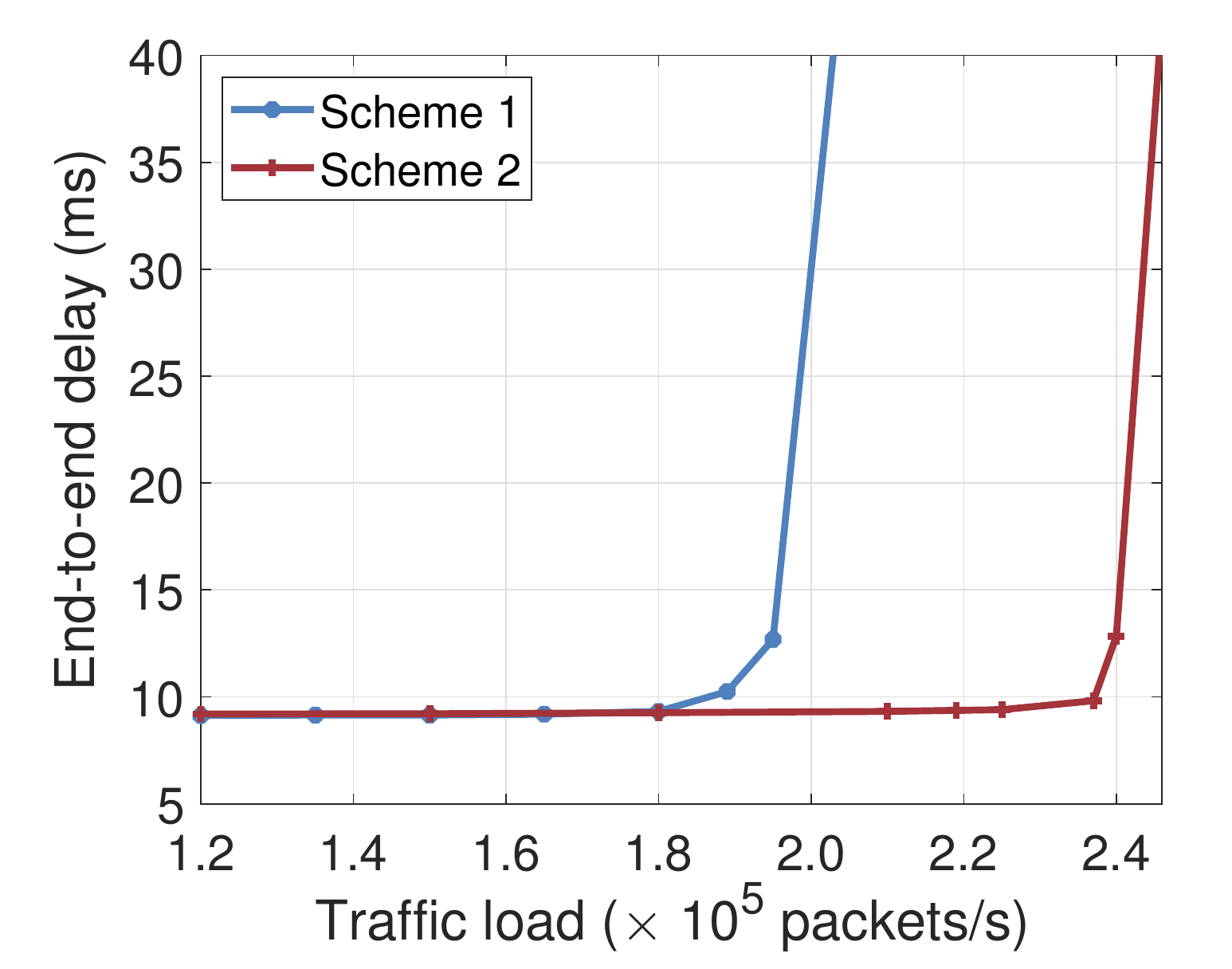}\label{fig:delay}}
	\caption{THz mesh network performance.}
\end{figure}

Based on well-maintained links and topologies, link layer functions such as resource allocation, power allocation, and interference management, and physical layer peculiarities in THz communications including the frequency- and distance-selective property need to be jointly considered to improve performance of THz mesh networking.
To illustrate the effectiveness of MAC/Phy cross-layer optimization, a THz mesh network consisting of $14$ mesh base stations is studied via simulations, in which mesh node 1 is the root base station providing a gateway to the Internet, while other mesh nodes are small base stations located in an area of 240m$\times$180m.
The key parameters in simulations include frequency at 460 GHz, bandwidth over 70 GHz, sub-band bandwidth over 5 GHz, maximum transmit power of 2 W, slot size equal to 0.125 ms, modulation mode of 4-PPM, noise figure at 7 dB, 64 antennas, PHY processing delay of 0.09 ms, and 3 ms RLC processing delay.
The root base station generates traffic flows to all small base stations, while each small base station only generates a traffic flow to the root base station, following a Poisson distribution. 
Two MAC/Phy cross-layer optimization schemes are evaluated: scheme 1 considers joint scheduling of time slots, sub-bands, and power, while scheme 2 takes into account an additional dimension, i.e., frequency- and distance-selectivity. The performance results of these two schemes are compared in Fig.~\ref{fig:thru} and~\ref{fig:delay}, in which the center frequency at 0.46~THz, occupation of fourteen 5~GHz-wide sub-bands, and the packet size of $9000$ bytes are considered.
At a regime of low traffic load, two schemes achieve comparable performance. However, as the traffic load grows beyond $1.9\times10^5$ packets/s, the average throughput of scheme 2 is more than 25$\%$ higher than that of scheme 1. Correspondingly, the average end-to-end delay of a traffic flow increases abruptly at a much lower traffic load in scheme 1. The much higher performance achieved by scheme 2 is attributed to the additional dimension of resource allocation in THz networking, i.e., frequency- and distance-selectivity.





In THz mesh networking, cross-layer design between the link layer and the physical layer is necessary to ensure QoS requirements at both the link level and the end-to-end level. For example, beamforming patterns can be dynamically controlled at the MAC layer to increase the flexibility of resource allocation and interference management for multihop mesh networking.
For another example, to achieve a graceful tradeoff between efficiency and reliability (as well as latency) under variable channel conditions, it is critical to take into account hybrid ARQ (particularly type-II).

\subsection{Routing and Higher Layer Protocols for THz Mesh Networking}

In THz mesh networking, new challenges appear at the higher layers of the protocol stack. 
At the network layer, new routing mechanisms to overcome LoS blockage need to take into account the availability of relaying schemes, including the classic active relaying nodes, passive dielectric mirrors and novel RIS (as discussed in Sec.~\ref{sec:ris}), which can intelligently direct the signal towards the final destination.
Furthermore, it is significant to explore new routing performance metrics that take into consideration the channel molecular composition along with its effect on the available distance-dependent bandwidth.
Moreover, since routing is closely related to MAC and physical layer functions, joint design of a routing algorithm and underlying protocols in the MAC and physical layers is of great importance. 
As one step further, a routing protocol that utilizes  machine and deep learning can be developed to dynamically adapt routes in real-time, without severely degrading the overall network throughput.
It is expected that IPv6 should be sufficient for classical macroscale communications when addressing is concerned, but for nanoscale applications, more advanced addressing paradigms, such as hierarchical-fashioned, need to be explored.

With the realization of wireless ultra-high-speed links, from multi-Gb/s to Tb/s, the aggregated traffic flowing through the network dramatically increases. At the transport layer, heavy burden is imposed on both congestion control and end-to-end reliable transport. The TCP congestion control window mechanism should be substantially revised, in order to handle the traffic dynamics in the 6G THz communication networks.
Considering the extremely high data rates in THz communications and super-low end-to-end latency requirements by many applications in 6G networks, simplifying interactions across protocol stacks becomes really important. This is particularly necessary for THz mesh networking. A potentially promising approach is to merge the network and transport layers with the link layer. To this end, layer-2 routing and transport protocols need to be developed for THz mesh networking. These protocols do not only improve the efficiency of protocol stack, but also enable natural operation of joint optimization among MAC, routing, and transport functions.


\section{Conclusion}
\label{sec:conclusion}
The THz band (0.1-10 THz) is envisioned as a key technology for 6G wireless systems, to fulfill the needs of extremely high data rates and super-low end-to-end latency in the next decade. 
This promising research direction has attracted increasing global attention and standardization efforts.
In this article, we presented the key THz channel features and compare them with microwave and FSO systems. Moreover, recent advancements in wireless device technologies are surveyed. 
We further investigate the impact and guidelines of THz technologies on 6G wireless communication and networking. 
This paper serves as a roadmap for the realization of this new frontier in 6G wireless communications.



%



\ifCLASSOPTIONcaptionsoff
  \newpage
\fi



%

\balance

\bibliography{./Bibliography/References_ComMag,./Bibliography/CH_bib}
\bibliographystyle{IEEEtran}

\end{document}